# Do Large Language Models Favour Any Research Topics?

Mike Thelwall, University of Sheffield, UK.

Large Language Models (LLMs) can estimate the quality of published journal articles, potentially supporting human assessment when evaluations are needed. Whilst there are reasons to believe that LLMs may have biases in this role, there is no statistically strong evidence yet. The current article addresses this gap with an exploration of the types of articles that attract high or low LLM scores in 73,489 articles from 15 health and life sciences journals. Based on comparing the words in the titles and abstracts of higher and lower scoring articles for two LLMs in various ways, the results suggest that topics favoured by GPT-OSS-120B include viruses, genes and cells and its disfavoured topics include surveys, patients and students. It is not clear whether these patterns reflect underlying quality differences or AI biases, however. The same method found systematic differences between the topics favoured by GPT-OSS-120B and Gemma 3 27B, such as Gemma 3 27B giving relatively higher scores for machine learning research, proving that at least one of the two LLMs has AI bias. Finally, comparing the scores for full-text articles compared to scores for titles and abstracts also finds differences for both LLMs, showing that they both can exhibit AI bias for at least one of these two input types, and probably both. Overall, the results show that it is important to consider LLM biases when deciding whether to use them for research evaluation tasks.

## 1 Introduction

Post-publication research quality evaluation is an increasingly common task for senior researchers, whether as part of appointment, tenure, and promotion procedures, for applicant evaluation in career-based funding applications, for internal departmental quality monitoring, or as part of nationally organised research assessment processes, such as the UK's Research Excellence Framework (REF). In all these cases, experts may lack the time or the specialist expertise to complete a thorough and effective review. Under this pressure, they may resort to shortcuts like Journal Impact Factors (JIFs), journal reputation, author reputation, or citation data to help. Large Language Model (LLM) judgement (Liang et al., 2024; Zheng et al., 2023) provides a new alternative with a proven ability to score journal articles in a way that correlates positively (weakly to moderately depending on the field) with expert judgements, if careful prompting strategies are used (Huang et al., 2025; Thelwall, 2026a; Zhu et al., 2026). Nevertheless, whilst the limitations of JIFs (Seglen, 1997; Waltman & Traag, 2021) and the biases of citation data are well known (Nørgaard et al., 2025; Ray et al., 2024), LLM scores have received little critical scrutiny. Bias here is different from inaccuracy and refers to any systematic deviation from the "correct" quality score, which may be theorised to be the mean or modal score given by a large group of human experts. Thus, an LLM bias would occur if it tended to give relatively high or low scores to one or more topics compared to human experts.

As background to the current investigation, there are some systematic topic differences in the quality scores assigned by experts to published journal articles. One study analysed the frequency of the words used in titles and abstracts to identify which associate with higher or lower expert scores, using 122,331 journal articles published 2014–2020 from the UK REF2021 national evaluation. It found that in all 34 fields, some terms associated with higher or lower expert scores. In particular, the first-person writing style associated with higher scores, perhaps because of its use by successful or medical researchers, and education and qualitative research seemed to attract lower scores (Thelwall et al., 2023). It is perhaps a philosophical question whether the lower expert scores for qualitative research represent a human bias.

Artificial Intelligence (AI) is known to be susceptible to bias by design or through its training data (Hanna et al., 2025; Mehrabi et al., 2021). This is also true for some and perhaps all LLMs (Lin et al., 2025; Wan et al., 2023), which may learn human biases from the human-authored texts used to train them, the reinforcement learning stage used to help generative AI LLMs respond appropriately to prompts (Casper et al., 2023; Ouyang et al., 2022), or any deliberate bias added by the creators for political reasons or for risk minimisation (Bai et al., 2022; Mu et al., 2024). For example, when scoring texts, LLMs are influenced by minor prompt wording variations, and may give higher scores to more predictable texts (Stureborg et al., 2024; Thelwall, 2026b; see also: Li et al., 2026). Moreover, the degree of bias varies between LLMs and tasks (Bavaresco et al., 2025).

For research assessment, it is known that technology can introduce bias (Thelwall & Kousha, 2025) and that title style can influence peer review scores (Du, 2025) and there are suggestions, but not proof, that LLMs may have biases for the task of research quality evaluation. Many papers have mentioned indirect evidence of bias, although it was not their focus. These have all been based on research quality scores assigned by LLMs (usually ChatGPT-4o mini) to published journal articles based on their titles and abstracts rather than their full texts. For clinical medicine, a comparison between ChatGPT-4o mini research quality scores and the authors' departmental average research quality scores (as a proxy for article-level scores) found that this LLM gave relatively higher scores to theoretical biology research (e.g., genetics) compared to clinical studies with patients, perhaps because biology research involves explicit originality, whereas health research can be conservative and focus on societal benefits and extreme robustness, therefore scoring on only two of the three dimensions. More clinical research also tended to attract relatively low ChatGPT-4o mini scores for its citation rates (Thelwall et al., 2025). An investigation of tourism research found that research with theories and advanced statistics scored higher and research with surveys scored lower with ChatGPT-4o mini: whilst these trends aligned with the articles' citation rates, they were not compared to human scores (Thelwall & Nunkoo, 2025). A similar study of library and information science journal articles found broadly similar results, with research about students and libraries also tending to get lower ChatGPT-4o mini scores (Thelwall, 2025a). One previous study directly addressing LLM biases has attempted a large-scale quantitative investigation of potential biases with a regression analysis of LLM scores against author country, a JIF variant, and citation rates. It found that more cited articles in higher impact journals with authors from large Western countries tend to get higher scores, although the associations varied between fields (Thelwall & Kurt, 2025). This study used only titles and abstracts as inputs, and it is not known whether there would be different associations for full-text inputs, or whether the results reflected LLM biases. Finally, one study has shown that paradigm bias can be injected into LLM scores if prompts are not worded neutrally (Thelwall et al., 2026).

Although the above review has found indirect evidence of LLM scoring bias, it has several gaps. First, it has not investigated *within*-journal biases, which is a limitation because all the patterns above could be caused by journal style differences in title or abstract text (e.g., length, structured headings, journal content guidelines or informal genres). Journals may be hypothesised to publish similar quality research, so systematic differences in LLM scores within a journal would be a new aspect from which to check for potential bias. Second, whilst clinical medicine has been investigated, the broader health and life sciences have not. This is an important gap since bias in the evaluation of research influencing life seems to be particularly problematic. Third, all studies have used scores derived from titles and abstracts, whereas full-text evaluations are more natural despite not giving results associating more strongly with human expert scores (Thelwall, 2025b). The following research questions address these three gaps.

- RQ1: Do any types of health and life science journal articles receive higher scores from LLMs both within and across journals?
- RQ2: Do the types of health and life science journal articles, if any, that receive higher scores from LLMs both within and across journals differ between LLMs?
- RQ3: Do the types of health and life science journal articles, if any, that receive higher scores from LLMs both within and across journals differ based on whether the LLM is fed with the full text or the title and abstract?

# 2 Methods

The research design was to:

1. Obtain full-text journal articles from a selection of large health and life sciences journals.
2. Score the articles for research quality with two different LLMs using both title/abstract and full-text inputs.
3. Use word association chi-squared tests with family-wise error rate correction to test for word frequency differences between higher and lower scoring subsets to address the three research questions.

## *2.1 Data: articles and factors*

A single journal publisher was selected as the source of all articles to remove the possibility that publisher-level factors influence the results of some research questions. The Swiss publisher MDPI was selected since its journals can be large, cover a wide range of life sciences fields, and are available in XML format for clean conversion to plain text for LLM processing. The fifteen largest life sciences journals were selected from its portfolio. Large journals were needed to give statistical power to the tests. For each one a random number generator in Python was used to select a random sample of 5,000 articles from the years 2020 to 2023 for each journal, or all articles if there were fewer. Multiple years were included to allow large sample sizes. The years 2020 to 2023 were selected to largely predate the LLM research era, which might affect the results through LLM-authored language in articles.

For each article, the title, abstract, and main body text (without references), were extracted from the XML. The title and abstract were combined for the title/abstract inputs, and all three elements were combined for the full-text inputs.

### 2.2 LLMs: Models and prompts

Although most prior research has used ChatGPT for testing, open-weight LLMs have similar performance, with Google's Gemma 3 27B being particularly effective (Thelwall & Mohammadi, 2026). Because of the large input volume, open-weight LLMs were used instead of cloud-based pay versions, with Gemma 3 27B being selected as the obvious choice. The newer and larger LLM GPT-OSS-120B from OpenAI was selected for comparison as the leading open-weight offering from one of the major LLM companies. The results foreground GPT-OSS-120B since this is the newer model.

The system and user prompts were the same used previously (Thelwall, 2026a). The system prompt defined the task using the official wording of REF2021 for health and life sciences research. This defines research quality in terms of rigour, originality and significance and gives the REF four-point scoring scheme from 1* (nationally relevant) to 4* (world leading). The user prompt requested separate reports for rigour, originality, and significance, and overall, each preceded by a short report. Either the title and abstract or the title, abstract and full text followed the user prompt below (Thelwall, 2026a).

> Score this article in the range 1* to 4*, including fractions. Use the following format:
> Rigour report:
> Rigour score:
> Originality report:
> Originality score:
> Significance report:
> Significance score:
> Overall report:
> Overall score:
> ###
> [article title and abstract <and full text>]

Each article was submitted 5 times for each prompt and the average of the five overall scores used as the final score for the article. Responses to identical prompts vary between runs and can yield different scores: averaging these scores gives better results in the sense of correlating more strongly with expert scores (Thelwall, 2026a).

### 2.3 Statistical Analysis

A word frequency test was used to identify systematic differences between higher and lower scoring articles. Although bias can be found by modelling techniques, such as regression (Thelwall & Kurt, 2025), this approach does not find unknown patterns. There are many text analysis methods that could be used, such as topic modelling, but few give safe statistical evidence of patterns. The method chosen was to check each word to see if it tended to occur disproportionately often in higher scoring or lower scoring texts. A 2 x 2 chi-squared test was used for this, which reports whether a term occurs in a statistically significantly different proportion of higher scoring texts to lower scoring texts, with the median score chosen as a cut-off (e.g., higher scoring vs. lower scoring and containing term vs. not containing term). For example, if the term "rigorous" occurred in 12% of the higher scoring half of the journal articles and 11% of the lower scoring half of the journal articles, then this chi-squared test would reveal whether the 1% difference was large enough to be unlikely to be due to chance.

Since this method entails a separate test for each word and there are many words, false positives are almost certain to occur. The Benjamini-Hochberg (Benjamini & Hochberg, 1995) procedure was used to prevent this. This substantially increases the requirements for a statistically significant result to prevent accidental positives from multiple tests, preserving the overall (family-wise) p value for the entire set of texts. This means that the results have some statistical rigour, but the likelihood of false negatives is increased by the Benjamini-Hochberg procedure as well as polysemy and synonyms. The root chi-squared test is imperfect, however, since its independence assumption can be violated by any cause of multiple similar high/low quality articles on the same topic, such as sets of articles by a single author, research team or temporarily important issue.

Variations of this chi-squared test were used to compare between models or inputs. For example, to find words associated with articles scored differently by the two models, the score difference was first calculated for each article, then the procedure above repeated for this score difference. This would find words occurring disproportionately often in articles with an above median score difference compared to a below median score difference.

### *2.4 Descriptive Analysis*

Although the word frequency chi-squared tests reported above can identify statistical evidence of score differences, they are less useful for identifying the patterns in the types of articles tending to attract higher or lower scores, as well as the magnitude of the score differences between them.

To address the first gap, cluster analysis was used to group articles into related sets and the average score of each set was calculated. The cluster analysis used k-means with a preset number of 25 clusters (a pragmatic choice to allow graphs), with the standard TF-IDF weighting metric applied to titles and abstracts, based on words, two-word phrases and three-word phrases combined. This is descriptive but not statistically robust because there are many types of cluster analysis and they give different outputs. Also, there are many different options that can also influence the results.

To address the second gap, for all one, two, or three word phrases occurring in at least 25 titles and abstracts, the mean score of the articles containing them was calculated, and the highest and lowest scoring terms and phrases extracted to illustrate the potential score magnitudes. This is not statistically robust because of the choice of minimum numbers of articles, and 95% confidence intervals calculated for this would not consider family-wise error rates from the many possible pairwise comparisons.

## 3 Results

Fifteen large life and health sciences journals were selected, all except one of which published at least 5000 articles during 2020-2023. These cover from health systems (*Healthcare*) and often large whole organism journals (e.g., *Animals*, *Plants*) to small scale analyses (e.g., *Molecules*, *Viruses*). Their scopes are relatively general for academic journals.

The average scores varied between LLM and input, with Gemma 3's full-text analysis producing the highest scores overall (2.944) and the lowest scores being produced by GPT-OSS-120B on titles/abstracts (2.456). Although the magnitude of the mean scores varies, the four sets of 15 journal average scores correlate strongly with each other, varying between r = 0.93 (GPT-OSS-120B titles/abstracts vs. Gemma 3 full texts) and r = 0.99 (GPT-OSS-120B titles/abstracts vs. GPT-OSS-120B full texts).

Table 1. The selected MDPI journals and the number of randomly selected articles selected from 2020-2023. The scores are the mean of 5 scores per article from the given LLM processing the stated input.

| **Journal** | **Articles** | **Abstract score** | | **Full-text score** | |
|---|---|---|---|---|---|
| | | **GPT** | **Gemma 3** | **GPT** | **Gemma 3** |
| Agronomy | 5000 | 2.447 | 2.668 | 2.579 | 2.993 |
| Animals | 5000 | 2.413 | 2.674 | 2.578 | 2.944 |
| Biomedicines | 5000 | 2.484 | 2.730 | 2.609 | 2.951 |
| Cancers | 5000 | 2.590 | 2.803 | 2.750 | 3.048 |
| Diagnostics | 5000 | 2.376 | 2.665 | 2.478 | 2.894 |
| Foods | 5000 | 2.414 | 2.650 | 2.592 | 2.960 |
| Healthcare | 5000 | 2.255 | 2.526 | 2.409 | 2.803 |
| International J. of Molecular Sciences (IJMS) | 5000 | 2.579 | 2.796 | 2.736 | 3.010 |
| Journal of Clinical Medicine (JCM) | 5000 | 2.390 | 2.644 | 2.525 | 2.863 |
| Medicina | 3489 | 2.209 | 2.457 | 2.346 | 2.671 |
| Microorganisms | 5000 | 2.582 | 2.770 | 2.748 | 3.034 |
| Molecules | 5000 | 2.437 | 2.653 | 2.566 | 2.873 |
| Nutrients | 5000 | 2.426 | 2.707 | 2.603 | 2.954 |
| Plants | 5000 | 2.466 | 2.693 | 2.629 | 2.981 |
| Viruses | 5000 | 2.689 | 2.839 | 2.855 | 3.095 |
| **Total** | **73489** | **2.456** | **2.690** | **2.605** | **2.944** |

### *3.1 Overall differences*

Many words occurred statistically significantly more often in the titles and abstracts from the 15 life science journals, according to GPT-OSS-120B (Table 2). These results suggest that there are some objects of study that generated above average scores from GPT-OSS-120B. These included the following, which align moderately well with the main clustering results below:

- Viruses (virus, viruses, viral, infection, host)
- Cells (cell, cells; associated with expression, cancer, associated with [*in*] *vitro*)
- Genes (gene, genes, expression)
- Genomics (genome, genomic; associated with genes, genetic)
- Mice (mice, mouse, murine, associated with [*in*] *vivo*)
- Proteins (protein, proteins, associated with cells, expression)

The non-topic, non-stylistic terms within the top term list include the following.

- Novel (novel; associated with therapeutic, cell, cells, target, discovery)
- Mechanisms (mechanisms, mechanism; associated with underlying, expression, molecular, cells)

Table 2. The 30 terms occurring most statistically significantly more often in high scoring (above median) than in low scoring (below median) articles across all 15 life science journals combined, using GPT-OSS-120B scores (average of 5 per article).

| Word | High score | Low score | Articles | Chisq. |
|---|---|---|---|---|
| we | 55.5% | 39.1% | 35109 | 1972.6 |
| here | 12.7% | 4.3% | 6405 | 1617.8 |
| virus | 8.7% | 3.2% | 4464 | 973 |
| novel | 12.0% | 6.1% | 6783 | 756 |
| viral | 6.3% | 2.2% | 3222 | 738.7 |
| cells | 18.1% | 11.0% | 10833 | 726.3 |
| gene | 13.6% | 7.4% | 7844 | 719.7 |
| viruses | 4.0% | 0.9% | 1838 | 712.5 |
| genome | 5.2% | 1.6% | 2590 | 705.5 |
| genes | 12.6% | 6.8% | 7242 | 697 |
| expression | 17.0% | 10.6% | 10264 | 622.4 |
| host | 4.7% | 1.5% | 2370 | 597.3 |
| rna | 4.7% | 1.6% | 2392 | 560.7 |
| sequencing | 6.7% | 3.0% | 3625 | 537.8 |
| cell | 17.5% | 11.5% | 10762 | 526.2 |
| identified | 17.3% | 11.4% | 10700 | 508.5 |
| mice | 6.2% | 2.8% | 3368 | 479.6 |
| replication | 2.6% | 0.7% | 1254 | 425.1 |
| infection | 9.7% | 5.7% | 5725 | 407.7 |
| previously | 6.0% | 3.0% | 3386 | 381.5 |
| mechanisms | 8.1% | 4.6% | 4737 | 379.8 |
| genomic | 3.3% | 1.2% | 1711 | 377.2 |
| highly | 8.1% | 4.6% | 4730 | 370.6 |
| proteins | 8.0% | 4.5% | 4667 | 358.3 |
| vivo | 5.6% | 2.8% | 3176 | 348.1 |
| genetic | 7.3% | 4.1% | 4235 | 345.9 |
| these | 35.4% | 29.0% | 23813 | 342.4 |
| strain | 4.5% | 2.0% | 2428 | 342.4 |
| mouse | 3.4% | 1.3% | 1800 | 339.1 |
| sequences | 3.3% | 1.3% | 1722 | 338.2 |

The terms found disproportionately often in below-median scoring articles had a very different pattern (Table 3). One set associated with structured abstract heading terms (e.g., objectives, materials, methods, conclusions, results, aim) and two terms associated with indirect language (was, were), contrasting with the direct language of “we” mentioned above. There were also several topics or methods:

- Questionnaires (questionnaire, associated with cross-sectional, participants, online, survey).
- Statistics (statistically, significant, p, ±, mean)
- Medical (medical; associated with records, hospital, care, patients)
- Hospital (hospital; associated with university, admitted, patients)
- Patients (patients, patient, associated with clinical, retrospective)

- Students (students, student, associated with undergraduate, nursing, education, questionnaire)

Table 3. The 30 terms occurring most statistically significantly more often in low scoring (below median) than in high scoring (above median) articles across all 15 life science journals combined, using GPT-OSS-120B scores (average of 5 per article).

| Word | Low score | High score | Articles | Chisq. |
|---|---|---|---|---|
| objectives | 8.0% | 2.5% | 3754 | 1151.4 |
| materials | 9.8% | 3.7% | 4853 | 1104.8 |
| methods | 26.8% | 16.8% | 15805 | 1073.8 |
| study | 72.7% | 61.3% | 48974 | 1063 |
| conclusions | 11.5% | 5.5% | 6141 | 868 |
| p | 28.3% | 20.1% | 17617 | 685.6 |
| were | 82.1% | 74.1% | 57211 | 670.3 |
| questionnaire | 5.3% | 1.8% | 2526 | 662.4 |
| aim | 15.9% | 9.7% | 9268 | 645.7 |
| results | 52.7% | 43.6% | 35176 | 603 |
| ± | 9.4% | 4.8% | 5126 | 580.7 |
| cross-sectional | 5.7% | 2.3% | 2885 | 570.4 |
| was | 80.3% | 72.9% | 56133 | 542.7 |
| group | 18.0% | 11.9% | 10848 | 540.9 |
| medical | 6.1% | 2.7% | 3171 | 495.6 |
| statistically | 4.9% | 1.9% | 2433 | 494.2 |
| significant | 26.1% | 19.3% | 16507 | 483.4 |
| hospital | 5.6% | 2.4% | 2886 | 482.2 |
| who | 12.4% | 7.6% | 7218 | 477.1 |
| retrospective | 7.3% | 3.7% | 3953 | 449.6 |
| groups | 15.9% | 10.8% | 9664 | 416.4 |
| mean | 9.5% | 5.6% | 5441 | 412.3 |
| university | 3.1% | 1.0% | 1462 | 410.6 |
| patients | 30.7% | 24.1% | 19997 | 391.9 |
| extract | 4.6% | 2.1% | 2400 | 383 |
| age | 13.1% | 8.6% | 7886 | 377.2 |
| evaluate | 14.1% | 9.5% | 8580 | 371.9 |
| students | 1.8% | 0.4% | 773 | 371.1 |
| antioxidant | 6.7% | 3.6% | 3682 | 369.9 |
| total | 23.1% | 17.5% | 14765 | 356.7 |

The strongest overall pattern across all 15 life sciences journals combined, in the sense of the highest chi-squared score in Table 2 or Table 3, was linguistic: using the first-person plural pronoun, sometimes with introductory phrases like, “Here we show…”. Despite this pattern, the word was common in both higher and lower scoring articles, with both having a similar score distribution (Figure 1) so the difference was overall rather than reflecting a cluster of high or low scoring articles.

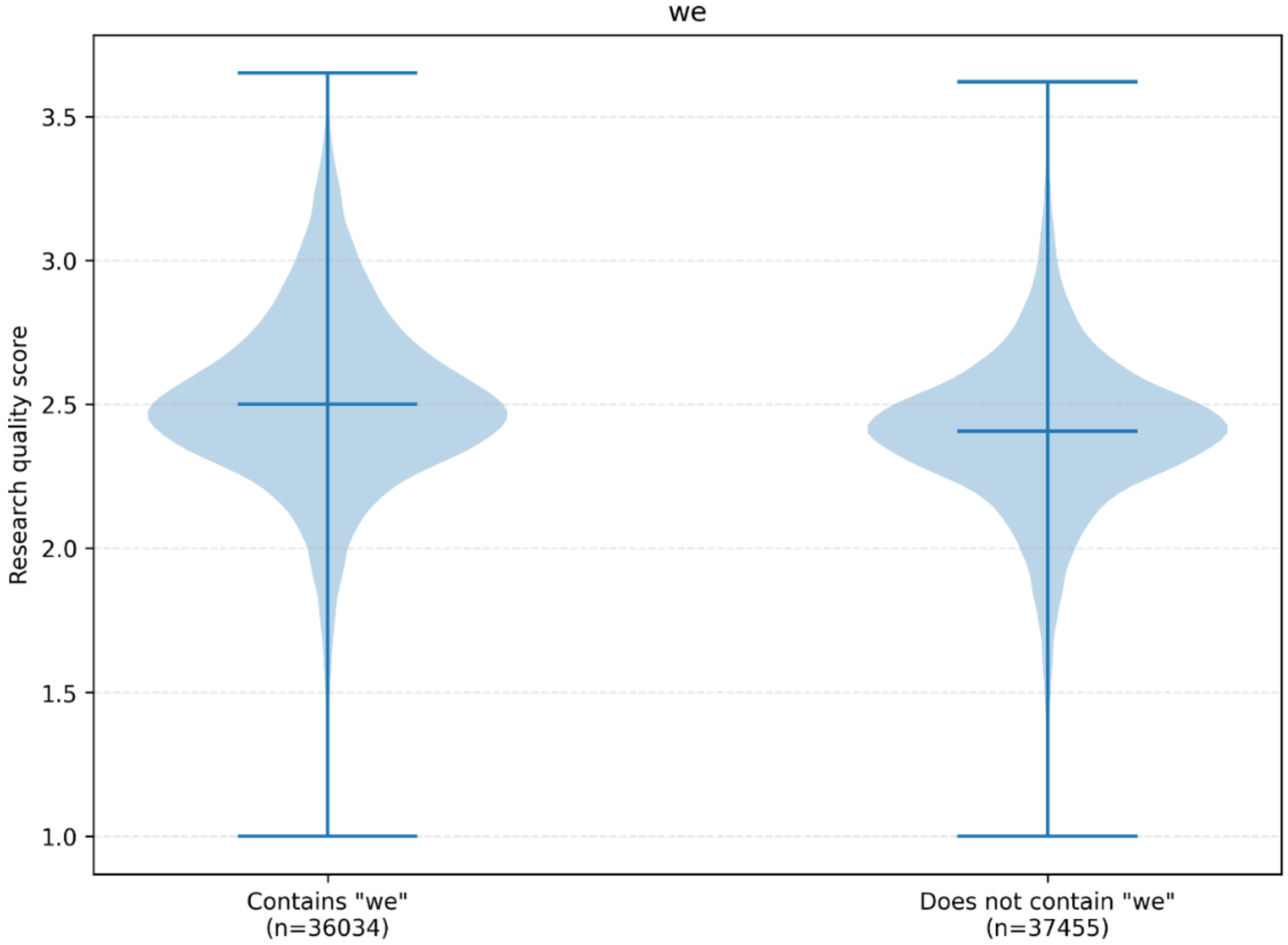

Figure 1. Violin plots of GPT-OSS-120B score frequencies for articles based on whether their title/ abstract contains the word “we” or not. Data: all 15 life science journal article titles and abstracts.

### 3.1.1 Descriptive analysis of overall differences

The descriptive analysis of the average scores for articles containing common words and phrases (2 or 3 words) found that many had high values (Figure 2). For example, articles with titles/abstracts containing “cryo-EM” (cryogenic electron microscopy) for imaging biomolecular structures at near to atomic resolution had the highest average score; capsids (virus protein shells) was second. There were also words and phrases that tended to be in article title/abstracts that attracted low scores, such as “dental students” and Iasi (city in Romania) (Figure 3). More generally, GPT-OSS-120B seemed to consider submissions from Romania and Saudi Arabia to be weaker than average, and the same for questionnaire-based articles (Figure 3).

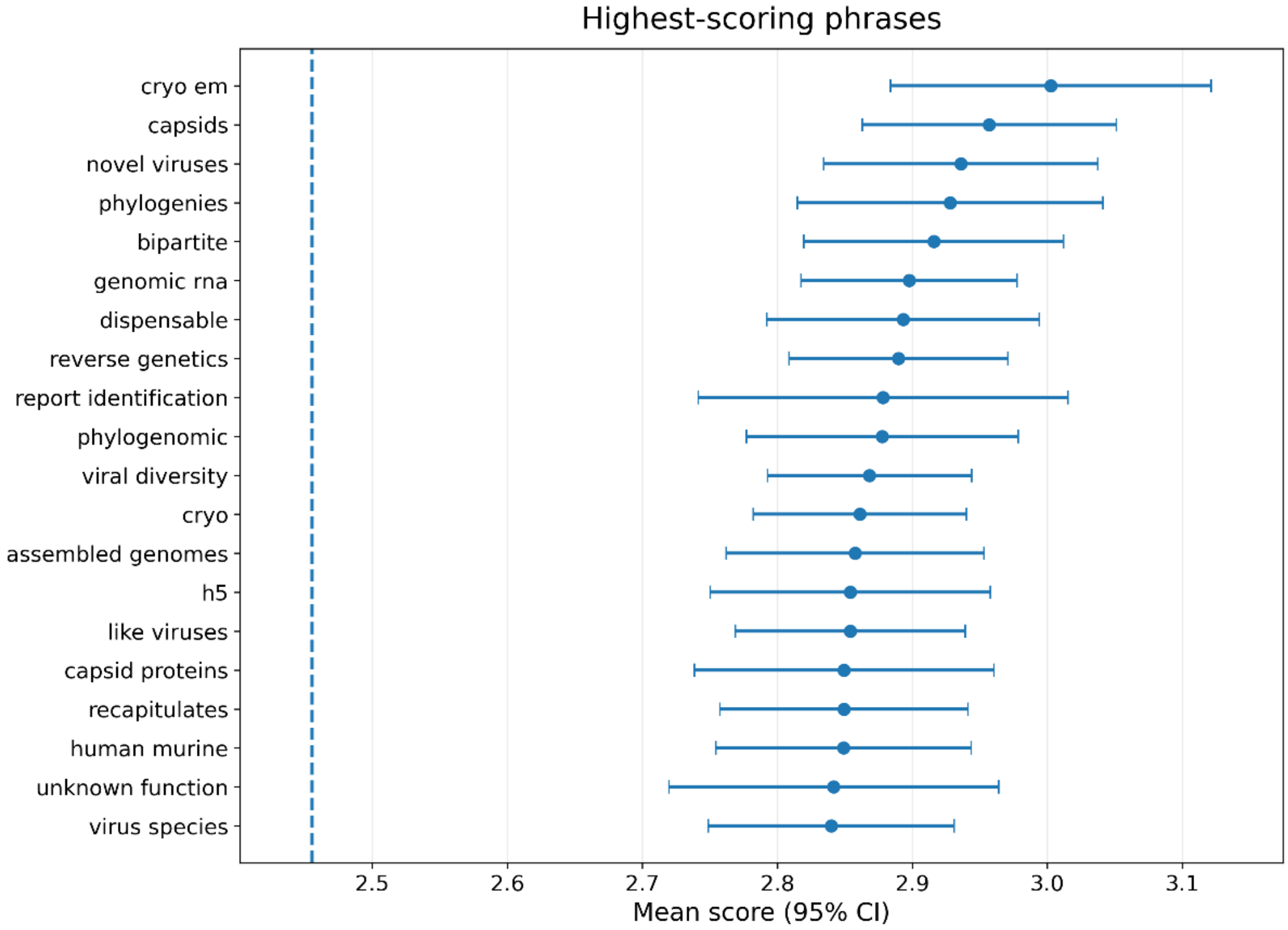

Figure 2. The 25 1-3 word phrases with the highest average GPT-OSS-120B scores for the article title/abstracts containing them across all 15 life science journals combined (n = 25 articles minimum). The dashed line is the overall mean.

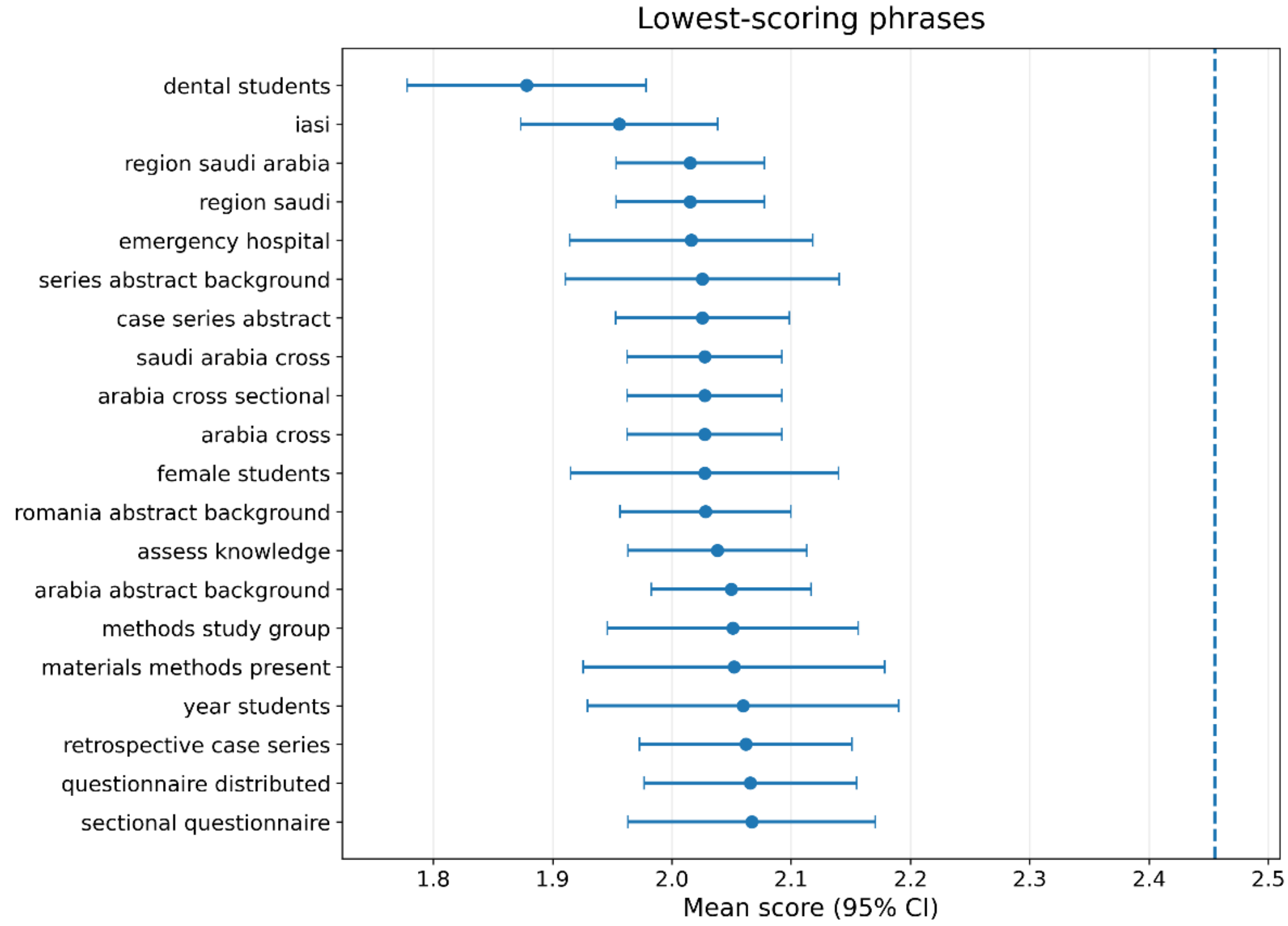

Figure 3. The 25 1-3 word phrases with the lowest average GPT-OSS-120B scores for the article title/abstracts containing them across all 15 life science journals combined (n = 25 articles minimum). The dashed line is the overall mean.

After clustering articles into 25 sets according to the textual similarity of their titles and abstracts, there are substantial differences between topics in terms of average scores (Figure 4). In particular, whilst studies of viruses seem to be the highest scoring, articles about Covid-19 and health seem to score the lowest. More generally, the most molecular and theoretical studies seem to score the highest, whereas human level studies (patients, women, health) seem to score below average.

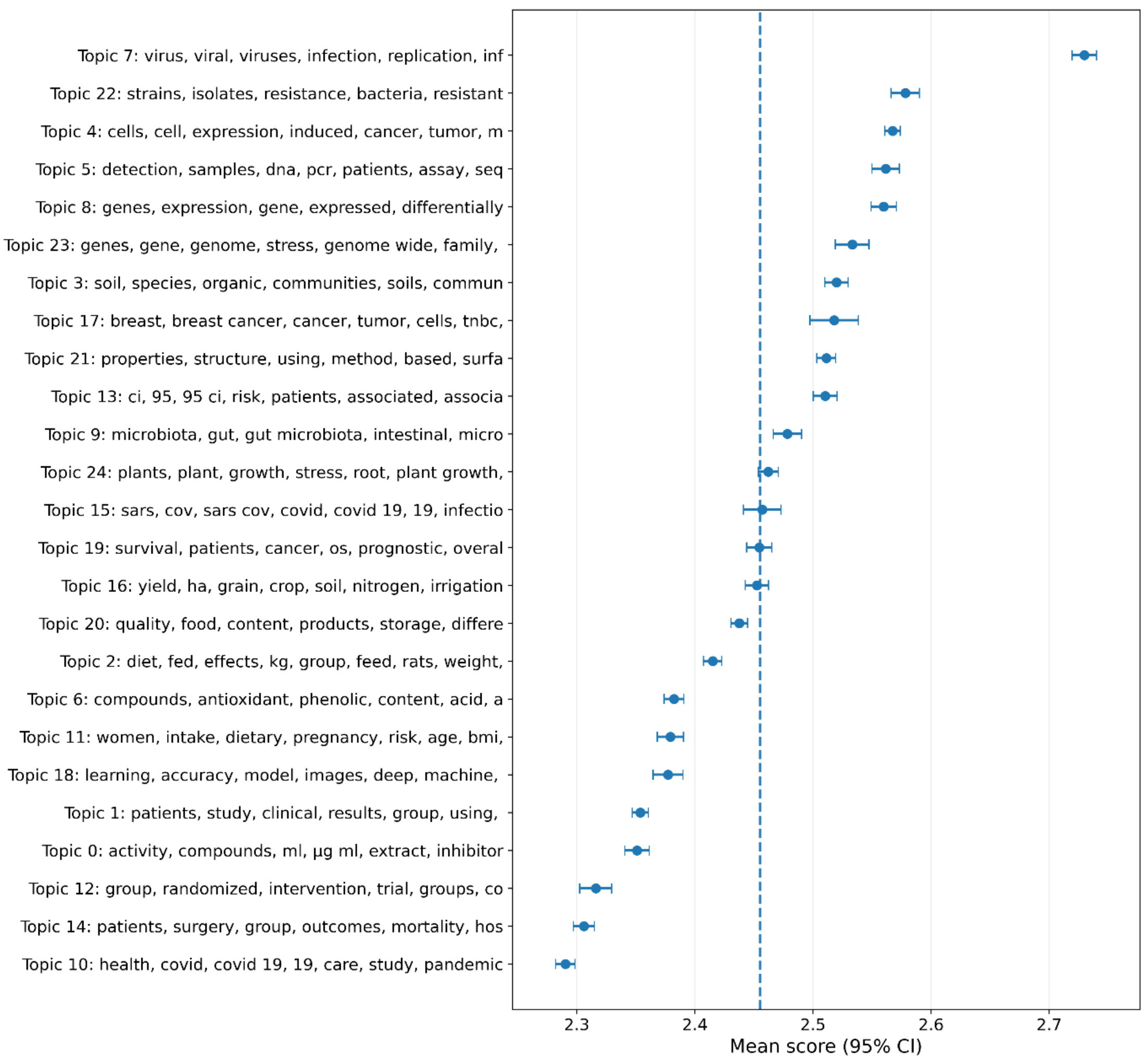

Figure 4. 25 article title/abstract clusters for all 15 life science journals combined and their mean GPT-OSS-120B title/abstract scores. Similarity was judged by TF-IDF based on 1-3 word phrases. The dashed line is the overall mean.

## *3.2 Within-journal differences*

The differences above could be caused by journal differences in average quality. Investigating within journals avoids this issue. For GPT-OSS-120B, many terms occurred disproportionately often in higher scoring articles (Table 4) or lower scoring articles (Table 5) for multiple journals. Both tables are led by stylistic and general terms, presumably since they could be used in many different journals. The tables confirm that the earlier results are not entirely due to journal differences since they occur within journals. They also show that specialist topics can be favoured by LLMs (e.g., the rockcress and model plant organism Arabidopsis).

Table 4. Terms statistically associating with **higher** GPT-OSS-120B quality scores for each of the 15 journals.

| Term | Journals |
|---|---|
| we | 11 |
| here | 8 |
| identified | 7 |
| associated | 6 |
| first, models, novel | 5 |
| adjusted, demonstrate, gene, genes, mice, model, mutant, national, previously, regression, show, vivo, within | 4 |
| by, ci, cohort, cox, genetic, genome, highly, insurance, is, logistic, multivariable, multivariate, mutants, our, protein, that | 3 |
| 0.95, a, across, activation, analysis, approach, between, breeding, but, cells, china, chinese, confidence, deletion, developed, development, disease, distinct, dna, encoding, evidence, extracellular, functional, genomes, genomic, growth, hazard, homeostasis, host, human, independently, known, loci, longitudinal, mouse, mutations, new, phenotype, population-based, populations, promotes, putative, qtl, ratio, ratios, regulates, related, report, resistance, risk, rna, sensitivity, sequencing, snps, species, specificity, strategies, structural, trait, transcriptional, transcriptomic, viruses, which, −/− | 2 |
| accessions, accurate, activated, activity, adaptation, additional, africa, agreement, allele, an, applied, arabidopsis, are, assay, assembly, assessed, association, associations, atrial, axis, binding, birth, bonds, burden, calculations, candidate, candidatus, capsid, carbon, catalysts, cell, cellular, challenge, characterized, chloroplast, chromatin, chromosome, clades, closely, cluster, complex, confirmatory, construct, containing, contributes, coupling, curve, database, defined, demonstrated, density, depletion, derived, described, detection, diets, divergent, diversity, driver, early, efficient, engineered, engineering, enriched, equation, essential, estimates, explore, expressing, expression, factor, family, feasibility, fed, fibroblasts, field, findings, formation, from, fully, genera, generated, grain, grapevine, gwas, hazards, health, heat, hosts, household, hr, hrs, identify, imaging, importantly, in-depth, incident, independent, induced, industrial, infants, infecting, infection, inhibition, innate, insights, integration, interacts, interval, interviews, involved, knockout, largely, lineage, lineages, linear, lines, long-term, loss, macrophage, major, maternal, mechanism, mechanisms, mediated, mediation, mendelian, microenvironment, milk, modeling, moderating, molecular, multicenter, murine, named, nationwide, nucleotide, odds, offspring, oncogenic, organoids, overcome, oxidation, panel, pathogen, pathway, patient-derived, phenotypes, photosynthetic, phylogenetic, phylogenomic, pluripotent, poorly, population, preclinical, prediction, predictive, pregnancy, profiling, prospectively, psychometric, qtls, qualitative, quantification, randomization, reaction, reactions, recently, receptor, recombinant, reduced, regions, registry, regulator, reliability, remains, reporter, representing, resistant, response, reveals, rice, rt-qpcr, segment, selection, semi-structured, set, severe, signaling, significantly, single, snp, spatial, still, stimulates, strategy, structure, suggesting, suppression, susceptible, systems, target, targeting, targets, testing, thematic, themes, thereby, these, together, transcription, transfer, uk, unclear, unique, upregulation, validity, viral, virion, virions, virus, vitro, weighted, welfare, wheat, work, xenograft, xenografts | 1 |

Table 5. Terms statistically associating with **lower** GPT-OSS-120B quality scores for each of the 15 journals.

| Term | Journals |
| --- | --- |
| study | 11 |
| aim | 10 |
| p, was, were, ± | 7 |
| patients, retrospective, treated | 6 |
| antioxidant, conclusions, evaluate, group, highest, hospital, no, results, significant, statistically | 5 |
| activity, effect, extract, pilot, treatment, value, values | 4 |
| activities, anti-inflammatory, antibacterial, anticancer, compounds, covid-19, cross-sectional, docking, effects, extracts, ic, included, inhibitory, it, medical, methods, parameters, properties, research, respectively, showed, single-center, students, there | 3 |
| =, according, after, against, aimed, antimicrobial, arabia, aureus, between, cancer, case, cases, cells, chemical, concentration, content, cytotoxic, cytotoxicity, diffusion, divided, dpph, evaluation, inhibition, january, levels, median, months, most, mtt, obtained, paper, period, phenolic, phytochemical, preliminary, questionnaire, received, romania, samples, sars-cov-2, saudi, scavenging, surgical, this, total, university, used, viability, vitro, who | 2 |
| abts, acetate, acid, acids, active, acute, admission, aeruginosa, age, anthropometric, antioxidants, apoptosis, assay, assess, asymptomatic, attitudes, bacillus, bacteroidetes, before, bioactive, biological, biomarkers, blood, blot, bmi, body, care, cell, center, centre, chromatography, classification, classify, clinical, cnn, cohort, coli, collected, color, compare, complications, composition, compound, concentrations, constituents, coronavirus, correct, county, days, death, december, dental, descriptive, determined, diagnosis, did, difference, disease, distributed, doctors, dried, due, early, eating, escherichia, ethanol, ethanolic, evaluated, experience, extraction, f1-score, fermented, first-line, flavonoids, follow-up, frap, fresh, fruits, gallic, googlenet, groups, habits, had, half, healthy, herbal, higher, hospitalization, hospitalized, hplc, hybrid, ifn-γ, igg, improvement, individuals, inflammatory, institution, kg, knowledge, l, lactic, lactobacillus, laparoscopic, leaves, line, lines, liquid, liver, material, mcf-7, medicinal, medicine, methanol, methanolic, mg, mg/ml, mic, microbiota, observational, on, oncological, one, online, oral, os, outcomes, overall, pandemic, participants, pattern, performed, pfs, phyla, physical, plantarum, polymorphism, polymorphisms, polyphenols, positive, postoperative, potent, potential, pre-trained, preoperative, probiotic, probiotics, procedure, proposed, purpose, radical, radiological, rate, rats, receiving, regarding, resection, respiratory, respondents, retrospectively, reviewed, rt-qpcr, rutin, safe, safety, significantly, silico, single, skin, spectrometry, spectroscopy, stage, staphylococcus, status, subjects, such, surgery, survival, syndrome, tannins, technique, tested, their, therapeutic, took, toxicity, traditional, underwent, various, vgg16, vitamin, western, women, years, µg/ml, μg/ml | 1 |

### 3.3 *Differences for the journal Cancers*

The single journal *Cancers* was selected for covering a high-profile human health topic to investigate the nature of the systematic topic score differences. For this, an additional 25 scores were obtained per article for a total of 30 per article to average for the final overall score. The purpose of this was to increase the quality of the score data to compensate for the smaller sample sizes. This dataset was only used for the graph (Figure 5). The difference between average cluster scores is statistically significant in many cases, with more theoretical and smaller scale research scoring higher and machine learning and patient research scoring lower.

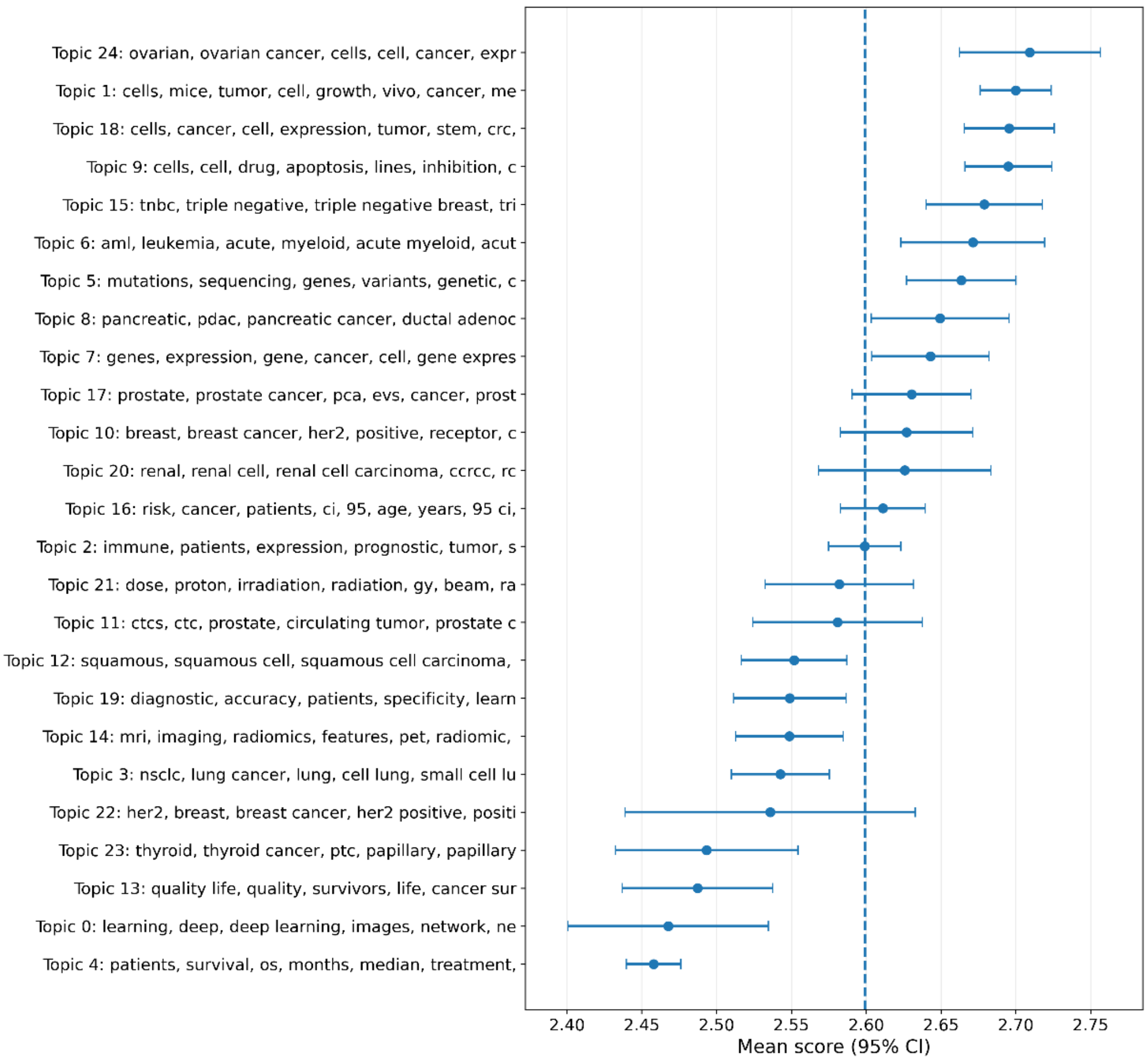

Figure 5. 25 article title/abstract clusters for *Cancers* and its mean GPT-OSS-120B title/abstract scores (average of 30 scores per article). Similarity was judged by TF-IDF based on 1-3 word phrases. The dashed line is the overall mean.

### 3.4 *RQ2: Comparison with Gemma 3 27B*

The same sets of terms were extracted for Gemma 3 27B to compare with GPT-OSS-120B. The results were similar overall, but not identical. For example, the terms most associated with high scores (Table 6) and most associated with low scores (Table 7) were largely the same for Gemma 3 27B and GPT-OSS-120B but with some differences in rank order. To illustrate this difference, the term “gut” was 140$^{th}$ most significant for high scoring articles for Gemma 3 27B but not in the top 1000 for GPT-OSS-120B.

Table 6. The top 10 terms occurring most statistically significantly more often in high scoring (above median) than in low scoring (below median) articles across all 15 life science journals combined, using Gemma 3 27B scores. Terms below the top 10 were selected to illustrate large rank differences

| Rank Gemma | Rank GPT | Word | High score | Low score | Articles | Chisq.* |
|---|---|---|---|---|---|---|
| 1 | 1 | we | 55.1% | 38.3% | 35109 | 2040.8 |
| 2 | 2 | here | 12.3% | 4.1% | 6405 | 1540.6 |
| 3 | 11 | expression | 17.4% | 9.5% | 10264 | 933.6 |
| 4 | 6 | cells | 18.1% | 10.4% | 10833 | 861.1 |
| 5 | 4 | novel | 11.9% | 5.8% | 6783 | 782.7 |
| 6 | 17 | mice | 6.4% | 2.2% | 3368 | 714.1 |
| 7 | 15 | cell | 17.6% | 10.9% | 10762 | 651.9 |
| 8 | 10 | genes | 12.3% | 6.7% | 7242 | 642 |
| 9 | 3 | virus | 8.0% | 3.6% | 4464 | 611.8 |
| 10 | 7 | gene | 13.1% | 7.5% | 7844 | 600.6 |
| 11 | 61 | model | 16.6% | 10.5% | 10249 | 550.2 |
| 14 | 67 | that | 71.1% | 63.3% | 49749 | 501.6 |
| 17 | 8 | viruses | 3.6% | 1.1% | 1838 | 446.6 |
| 21 | 9 | genome | 4.7% | 2.0% | 2590 | 413.9 |
| 27 | 121 | however | 27.5% | 21.4% | 18246 | 368.8 |
| 103 | 32 | infected | 3.8% | 2.1% | 2230 | 185.4 |
| 139 | 446 | validation | 3.0% | 1.6% | 1773 | 149.6 |
| 140 | - | gut | 2.5% | 1.3% | 1464 | 149 |
| 198 | 39 | phylogenetic | 2.6% | 1.4% | 1508 | 116.3 |
| - | 149 | surveillance | - | - | - | - |

- Outside the top 1000 terms.

*All scores are statistically significant at $p < 0.001$ after a Benjamini Hochberg correction.

Table 7. The top 10 terms occurring most statistically significantly more often in low scoring (below median) than in high scoring (above median) articles across all 15 life science journals combined, using Gemma 3 27B scores. Terms below the top 10 were selected to illustrate large rank differences

| Rank Gemma | Rank GPT | Word | Low score | High score | Articles | Chisq.* |
|---|---|---|---|---|---|---|
| 1 | 1 | objectives | 8.7% | 2.3% | 3754 | 1516.5 |
| 2 | 2 | materials | 10.5% | 3.6% | 4853 | 1409.2 |
| 3 | 3 | methods | 27.3% | 17.0% | 15805 | 1137.4 |
| 4 | 5 | conclusions | 12.2% | 5.4% | 6141 | 1083.2 |
| 5 | 4 | study | 72.1% | 62.4% | 48974 | 772.7 |
| 6 | 7 | were | 82.6% | 74.1% | 57211 | 761.1 |
| 7 | 8 | questionnaire | 5.5% | 1.8% | 2526 | 727.1 |
| 8 | 12 | cross-sectional | 6.1% | 2.3% | 2885 | 689.4 |
| 9 | 18 | hospital | 6.0% | 2.3% | 2886 | 658.5 |
| 10 | 9 | aim | 16.1% | 9.9% | 9268 | 645.4 |
| 11 | 26 | age | 14.0% | 8.2% | 7886 | 628.2 |
| 15 | 6 | p | 28.0% | 20.8% | 17617 | 515.5 |
| 17 | 60 | years | 16.4% | 10.8% | 9754 | 486 |
| 22 | 55 | survey | 4.9% | 2.2% | 2491 | 413.6 |
| 27 | 11 | ± | 9.0% | 5.4% | 5126 | 370.5 |
| 30 | 103 | january | 3.2% | 1.3% | 1574 | 306.1 |
| 33 | 14 | group | 17.3% | 12.8% | 10848 | 296.7 |
| 56 | 126 | gender | 2.3% | 1.0% | 1146 | 219.2 |
| 89 | - | prevalence | 5.2% | 3.4% | 3066 | 152.1 |
| - | 77 | training | - | - | - | - |

-Outside the top 1000 terms.

*All scores are statistically significant at $p < 0.001$ after a Benjamini Hochberg correction.

### 3.4.1 Descriptive analysis of Gemma 3 27B results

The highest scoring 1–3-word title/abstract phrases for Gemma 3 27B (Figure 6) also have a substantial but partial overlap with those for GPT-OSS-120B (Figure 2) and the same for the lowest-scoring 1-3-word title/abstract phrases (Figure 7). This confirms that similar topics, concepts, methods or wordings tend to associate with high and low scores in both cases. The topic hierarchy for Gemma 3 27B (Figure 8) is also similar to that for GPT-OSS-120B (Figure 3) suggesting that their results are likely to be broadly similar overall.

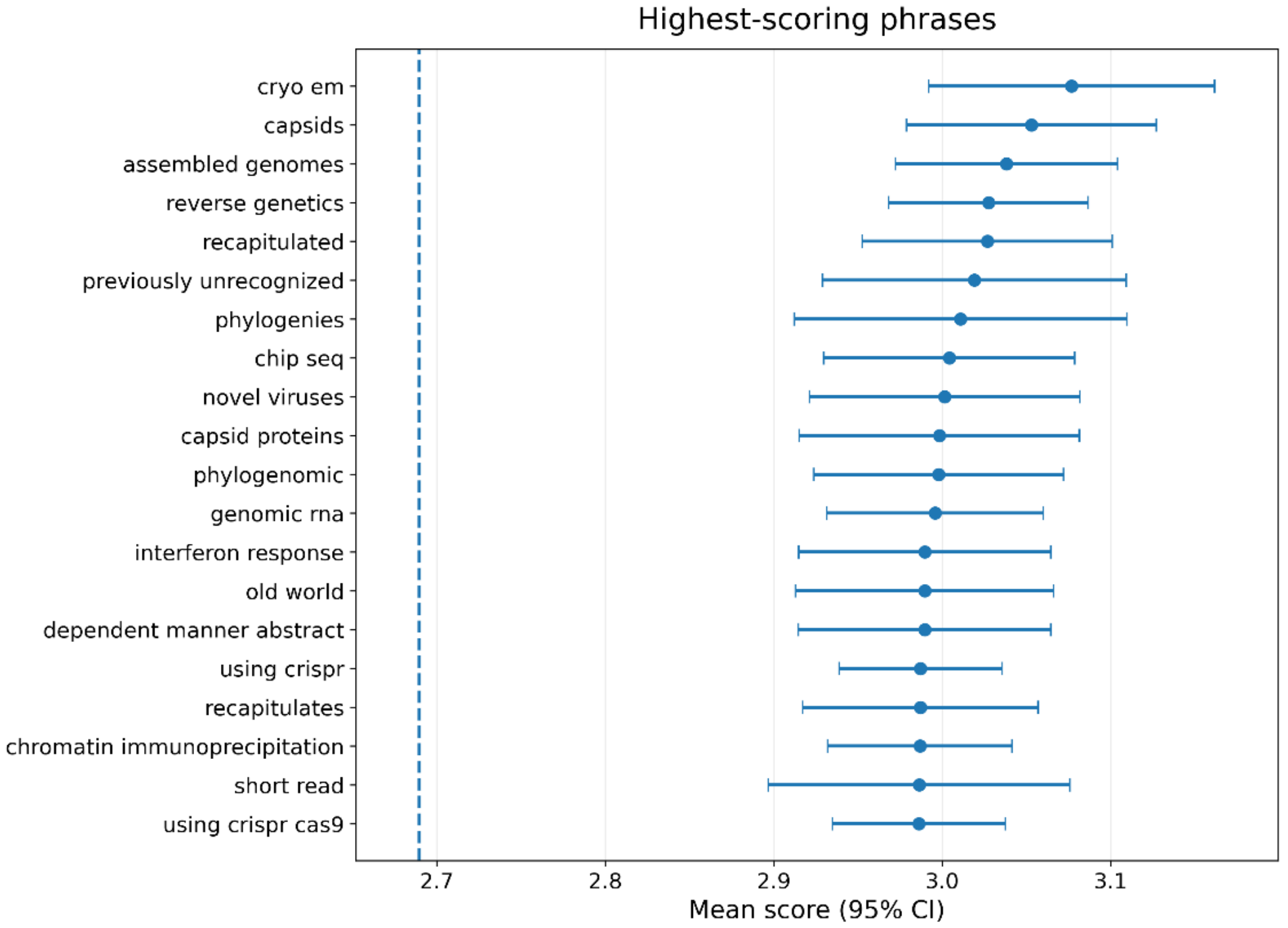

Figure 6. The 25 1–3-word phrases with the highest average Gemma 3 27B scores for the article title/abstracts containing them across all 15 life science journals combined (n = 25 articles minimum). The dashed line is the overall mean.

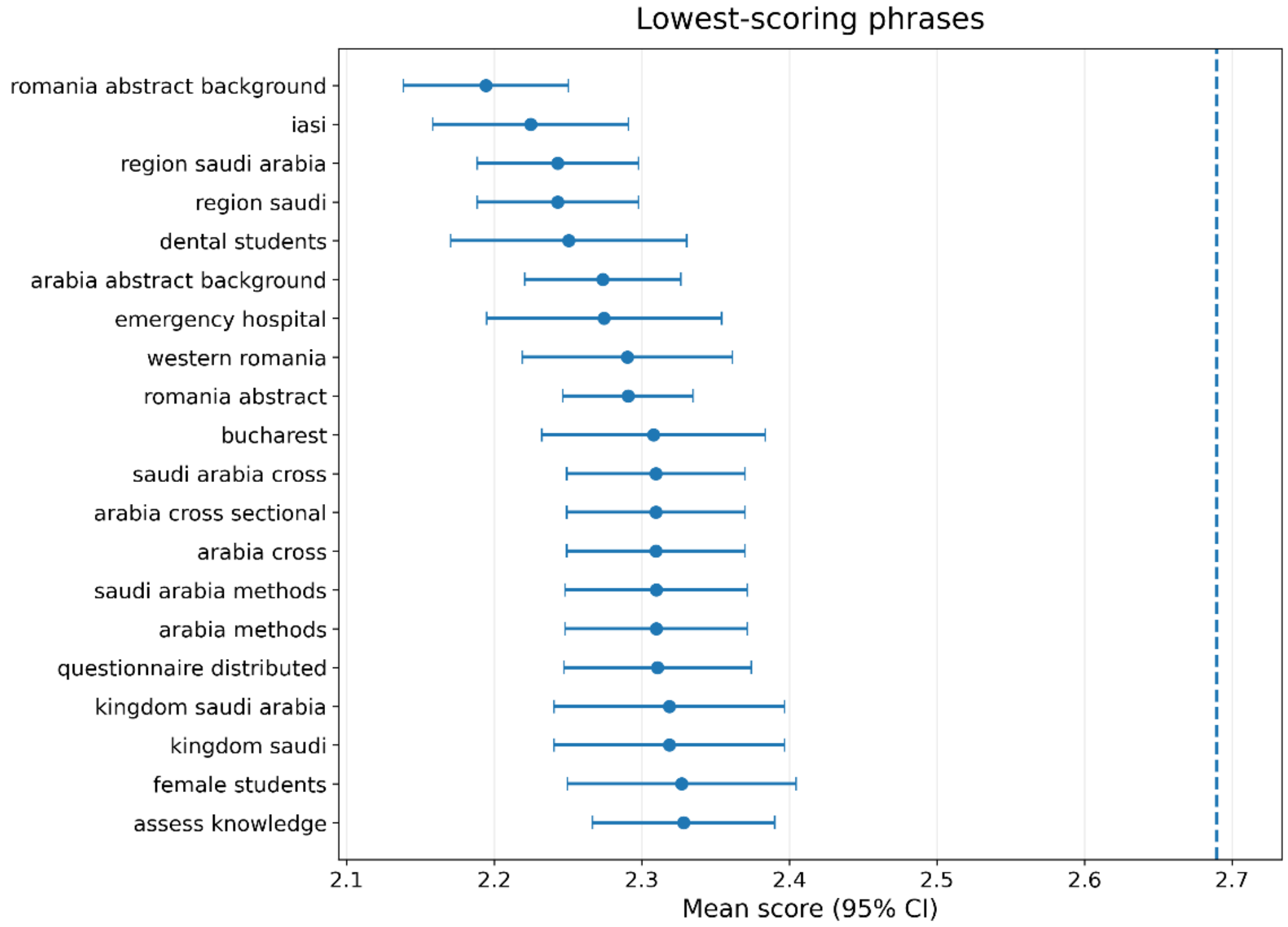

Figure 7. The 25 1–3-word phrases with the lowest average Gemma 3 27B scores for the article title/abstracts containing them across all 15 life science journals combined (n = 25 articles minimum). The dashed line is the overall mean.

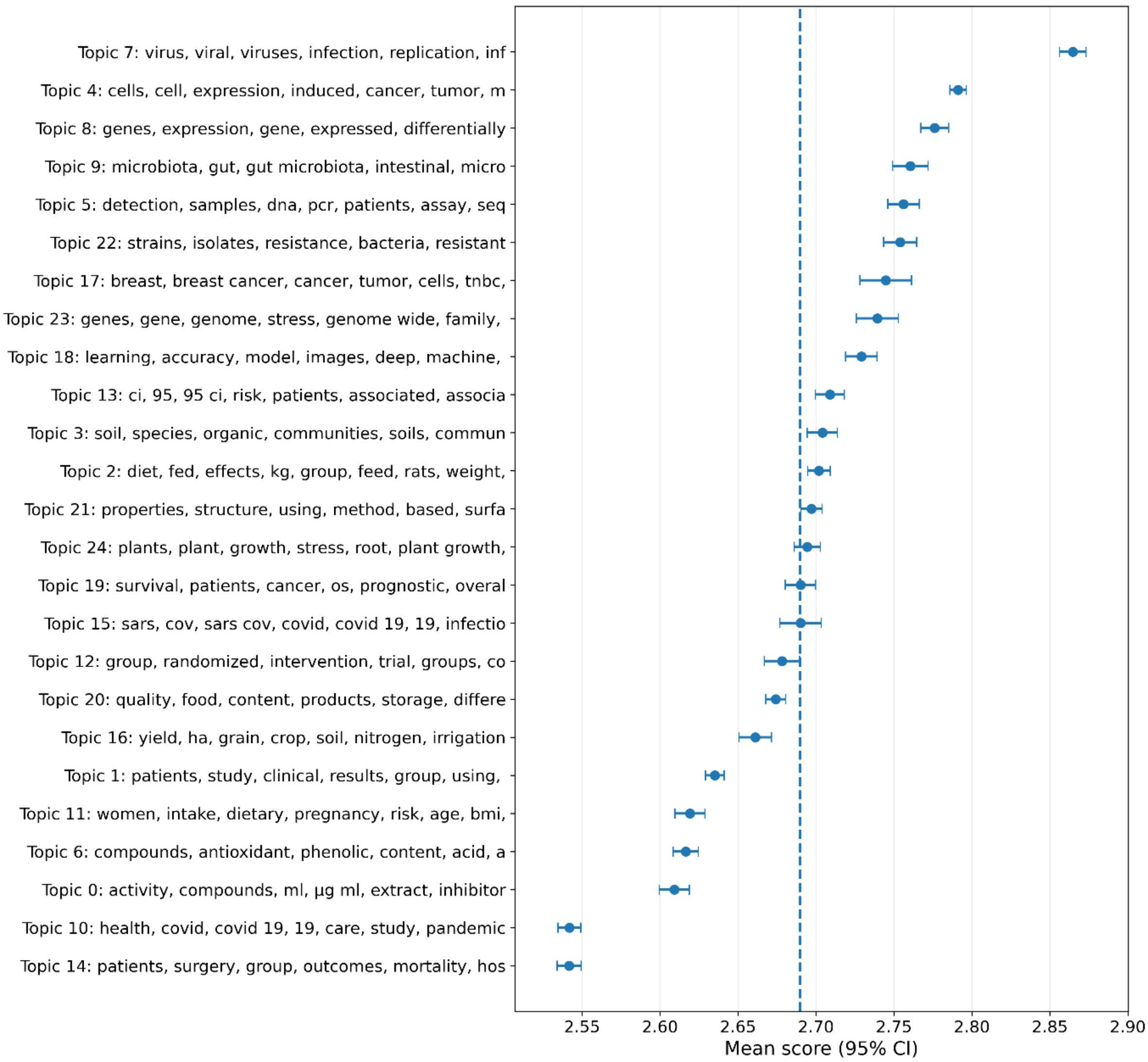

Figure 8. 25 article title/abstract clusters for all 15 life science journals combined and their mean Gemma 3 27B title/abstract scores. Similarity was judged by TF-IDF based on 1-3 word phrases. The dashed line is the overall mean.

### 3.4.2 Gemma 3 27B vs. GPT-OSS-120B

The difference between the two LLMs for the words most associating with high and low scores can be more directly investigated by subtracting the Gemma 3 27B score from the GPT-OSS-120B score and finding words that most associate with high score differences. The terms most associated with higher GPT-OSS-120B than Gemma 3 27B scores (Table 8) suggest that GPT-OSS-120B more strongly favours research about viruses and the genome. It also statistically favours titles/abstract mentioning China. The terms most associated with China in titles and abstracts are Chinese province names. A possible explanation for this is that either GPT-OSS-120B has a positive association with China or Gemma 3 27B has a negative association with China, perhaps for political reasons or topics unrelated to academic research for both possibilities. In contrast, the terms scoring higher for Gemma 3 27B compared to GPT-OSS-120B mostly seem related to machine learning (Table 9). A possible explanation is that GPT-OSS-120B's training is newer and traditional machine learning has been replaced to some extent with LLMs recently. Alternatively, Google might have access to more historical technical documentation or computing academic papers to train Gemma 3 with, perhaps giving it a relative strength in this technical area.

Table 8. The top 20 terms occurring most statistically significantly more often in high GPT-OSS-120B minus Gemma 3 27B scores (above median difference) than in low GPT-OSS-120B minus Gemma 3 27B scores (below median difference) articles across all 15 life science journals combined, using Gemma 3 27B scores.

| Word | Higher GPT-OSS-120B | Lower GPT-OSS-120B | Articles | Chisq.* |
|---|---|---|---|---|
| virus | 7.6% | 4.6% | 4464 | 299.9 |
| viruses | 3.4% | 1.6% | 1838 | 219.9 |
| viral | 5.4% | 3.4% | 3222 | 185.9 |
| genome | 4.4% | 2.6% | 2590 | 167.3 |
| diversity | 4.8% | 3.0% | 2871 | 163 |
| infection | 9.0% | 6.6% | 5725 | 153.1 |
| phylogenetic | 2.7% | 1.4% | 1508 | 138.6 |
| isolates | 2.9% | 1.7% | 1683 | 132.1 |
| prevalence | 5.0% | 3.4% | 3066 | 115.4 |
| china | 3.3% | 2.0% | 1929 | 113.2 |
| characterization | 5.0% | 3.4% | 3102 | 109.5 |
| host | 3.9% | 2.5% | 2370 | 108 |
| genetic | 6.7% | 4.9% | 4235 | 107 |
| surveillance | 1.8% | 0.9% | 1011 | 103.8 |
| genomes | 1.8% | 0.9% | 1006 | 103.6 |
| sequencing | 5.7% | 4.2% | 3625 | 96 |
| strains | 5.0% | 3.6% | 3159 | 94.9 |
| here | 9.7% | 7.7% | 6405 | 92.5 |
| national | 2.9% | 1.9% | 1770 | 88.7 |
| sequences | 2.9% | 1.8% | 1722 | 88.3 |

*All scores are statistically significant at $p < 0.001$ after a Benjamini Hochberg correction.

Table 9. The top 20 terms occurring most statistically significantly more often in low GPT-OSS-120B minus Gemma 3 27B scores (below median difference) than in high GPT-OSS-120B minus Gemma 3 27B scores (above median difference) articles across all 15 life science journals combined, using Gemma 3 27B scores.

| Word | Higher Gemma 3 27B | Lower Gemma 3 27B | Articles | Chisq.* |
|---|---|---|---|---|
| accuracy | 5.6% | 2.7% | 3081 | 382.1 |
| learning | 4.2% | 2.0% | 2265 | 305.4 |
| images | 3.7% | 1.8% | 2014 | 239.7 |
| training | 3.3% | 1.6% | 1788 | 232.3 |
| effects | 21.7% | 17.4% | 14390 | 212.7 |
| weeks | 5.5% | 3.3% | 3245 | 212.1 |
| pilot | 2.1% | 0.9% | 1097 | 200.6 |
| randomized | 3.3% | 1.7% | 1824 | 194.6 |
| supplementation | 3.1% | 1.5% | 1696 | 190 |
| image | 2.4% | 1.1% | 1267 | 174.9 |
| neural | 2.3% | 1.0% | 1213 | 170.4 |
| intervention | 3.9% | 2.2% | 2243 | 159.5 |
| machine | 2.5% | 1.2% | 1354 | 154.5 |
| group | 16.3% | 13.2% | 10848 | 145.9 |
| performance | 9.3% | 6.9% | 5948 | 145 |
| convolutional | 1.1% | 0.3% | 534 | 143.6 |
| diet | 5.4% | 3.6% | 3314 | 140.9 |
| body | 6.7% | 4.7% | 4217 | 135.3 |
| compared | 26.8% | 23.2% | 18396 | 126.3 |
| after | 24.3% | 20.8% | 16568 | 125.7 |

*All scores are statistically significant at $p < 0.001$ after a Benjamini Hochberg correction.

### 3.4.3 Gemma 3 27B vs. GPT-OSS-120B: Descriptive analysis

Although the ordering of the average scores of articles in the 25 clusters is broadly similar between the two models, there are systematic differences (Figure 9). For example, GPT-OSS-120B systematically scores virus research higher than Gemma 3 27B. In contrast, Gemma 3 27B systematically scores machine learning and randomised trial research higher than GPT-OSS-120B.

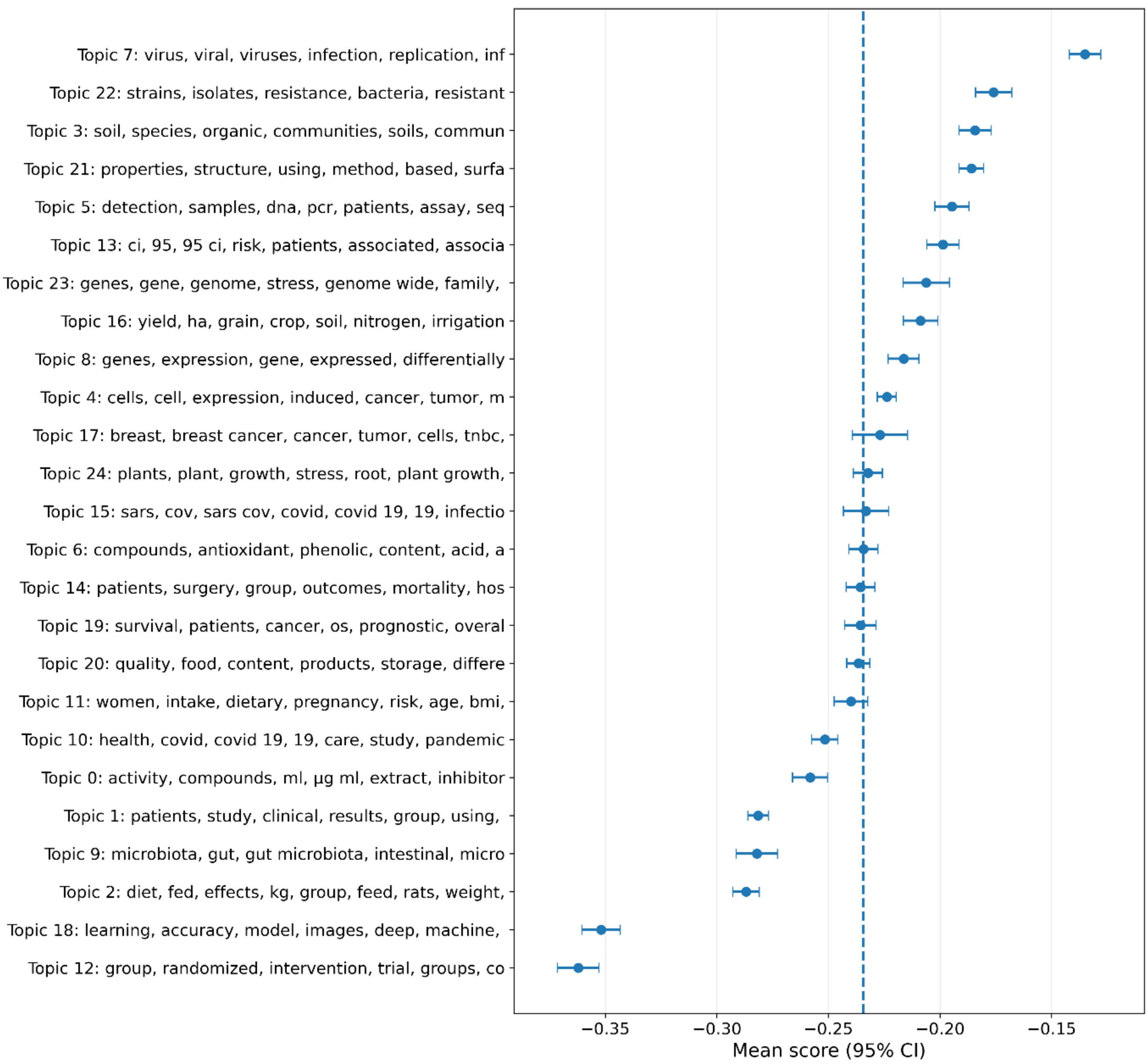

Figure 9. 25 article title/abstract clusters for all 15 life science journals combined and their mean GPT-OSS-120B title/abstract scores minus Gemma 3 27B title/abstract scores. Similarity was judged by TF-IDF based on 1-3 word phrases. The dashed line is the overall mean difference.

### *3.5 RQ3: Full text against title/abstract*

Many terms occurred disproportionately often in articles with higher scores for title/abstract than for full-text input (Table 10), or the other way around (Table 11) for GPT-OSS-120B and Gemma 3 27B (Table 12, Table 13). For example, articles mentioning patients were more likely to have a lower score from GPT-OSS-120B if scored on their full text than on their title and abstract (Table 10). This confirms that there are input biases for both GPT-OSS-120B and Gemma 3 27B in the sense of systematic score differences based on what the input is. The differences tend to be less substantial, as suggested by the lower chi-squared scores in the tables below compared to the tables above.

Table 10. The top 20 terms occurring most statistically significantly more often in scores for titles and abstracts minus scores for full texts (above median difference vs. below median difference) for articles across all 15 life science journals combined, using GPT-OSS-120B. Terms were only extracted from titles and abstracts.

| **Word** | **High abstract – full-text score** | **Low abstract – full-text score** | **Chisq.*** |
|---|---:|---:|---:|
| patients | 29.7% | 24.5% | 244.9 |
| significantly | 25.6% | 22.6% | 87.9 |
| analysis | 32.5% | 29.3% | 84.2 |
| accuracy | 4.8% | 3.5% | 75.7 |
| multivariate | 3.2% | 2.2% | 73.1 |
| underwent | 5.1% | 3.8% | 67.3 |
| prognostic | 3.6% | 2.6% | 65.2 |
| image | 2.1% | 1.3% | 57.2 |
| neural | 2.0% | 1.3% | 55.7 |
| treatment | 21.7% | 19.5% | 55.3 |
| images | 3.2% | 2.3% | 54.1 |
| classification | 3.3% | 2.4% | 52.5 |
| learning | 3.5% | 2.6% | 52.2 |
| regression | 6.9% | 5.6% | 51.2 |
| clinical | 16.5% | 14.6% | 50.5 |
| treated | 7.8% | 6.4% | 50.1 |
| results | 49.1% | 46.5% | 49.4 |
| curve | 2.4% | 1.6% | 48.9 |
| predicting | 2.4% | 1.7% | 47.3 |
| auc | 1.7% | 1.1% | 46.9 |

*All scores are statistically significant at $p < 0.001$ after a Benjamini Hochberg correction.

Table 11. The 18 terms occurring most statistically significantly more often in scores for full texts minus scores for titles and abstracts (above median difference vs. below median difference) for articles across all 15 life science journals combined, using GPT-OSS-120B. Terms were only extracted from titles and abstracts.

| Word | Low abstract – full-text score | High abstract – full-text score | Chisq.* |
|---|---|---|---|
| here | 9.7% | 7.9% | 74.3 |
| intake | 3.8% | 2.8% | 49 |
| species | 10.5% | 9.0% | 45.4 |
| food | 7.9% | 6.6% | 43.6 |
| viruses | 2.9% | 2.1% | 42.3 |
| composition | 7.0% | 5.8% | 42.2 |
| but | 21.7% | 19.9% | 35.8 |
| their | 29.0% | 27.1% | 33.5 |
| behavior | 3.4% | 2.7% | 33.3 |
| consumers | 1.3% | 0.9% | 33.2 |
| spain | 1.1% | 0.8% | 30 |
| not | 29.3% | 27.4% | 30 |
| flour | 0.7% | 0.4% | 28 |
| consumption | 3.6% | 2.9% | 26.5 |
| crossover | 0.4% | 0.2% | 26.2 |
| dietary | 4.6% | 3.8% | 26.2 |
| cross-sectional | 4.3% | 3.6% | 26.2 |
| allowed | 2.0% | 1.5% | 26 |

*All scores are statistically significant at $p < 0.05$ after a Benjamini Hochberg correction.

Table 12. The top 20 terms occurring most statistically significantly more often in scores for titles and abstracts minus scores for full texts (above median difference vs. below median difference) for articles across all 15 life science journals combined, using Gemma 3 27B. Terms were only extracted from titles and abstracts.

| Word | **High abstract – full-text score** | **Low abstract – full-text score** | Chisq.* |
|---|---|---|---|
| patients | 31.9% | 22.2% | 880.8 |
| we | 50.8% | 44.6% | 284.1 |
| = | 21.2% | 16.5% | 263.4 |
| p | 26.3% | 21.4% | 240.8 |
| group | 16.7% | 12.7% | 234.4 |
| expression | 15.8% | 12.0% | 226.6 |
| cells | 16.6% | 12.8% | 211.5 |
| blood | 8.7% | 5.9% | 207.6 |
| pilot | 2.1% | 0.8% | 193.6 |
| significantly | 26.2% | 21.9% | 185.7 |
| disease | 17.9% | 14.2% | 183.6 |
| after | 24.5% | 20.4% | 178 |
| whether | 7.9% | 5.5% | 171.6 |
| clinical | 17.3% | 13.8% | 170.7 |
| cell | 16.3% | 12.9% | 168.7 |
| controls | 4.7% | 2.9% | 167.4 |
| follow-up | 5.0% | 3.1% | 158.7 |
| not | 30.3% | 26.1% | 157 |
| underwent | 5.4% | 3.5% | 146.7 |
| retrospective | 6.3% | 4.4% | 137.2 |

*All scores are statistically significant at $p < 0.001$ after a Benjamini Hochberg correction.

Table 13. The 20 terms occurring most statistically significantly more often in scores for full texts minus scores for titles and abstracts (above median difference vs. below median difference) for articles across all 15 life science journals combined, using Gemma 3 27B. Terms were only extracted from titles and abstracts.

| Word | Low abstract – full-text score | High abstract – full-text score | Chisq.* |
|---|---|---|---|
| content | 12.2% | 8.1% | 336.8 |
| soil | 5.3% | 2.8% | 277.8 |
| yield | 6.7% | 4.2% | 216.7 |
| highest | 8.6% | 6.0% | 182.1 |
| quality | 11.5% | 8.5% | 181 |
| plant | 9.5% | 6.9% | 166.4 |
| different | 26.7% | 22.7% | 164.3 |
| crop | 4.2% | 2.5% | 155.1 |
| l | 7.6% | 5.4% | 151.4 |
| climate | 2.0% | 0.9% | 138.4 |
| nitrogen | 3.2% | 1.9% | 126.4 |
| compounds | 8.0% | 6.0% | 120 |
| conditions | 11.7% | 9.2% | 119.5 |
| organic | 3.8% | 2.4% | 115.8 |
| water | 6.8% | 5.0% | 112.3 |
| plants | 7.1% | 5.2% | 111.1 |
| agricultural | 2.5% | 1.5% | 103 |
| leaf | 3.4% | 2.2% | 102.2 |
| fertilizer | 1.4% | 0.6% | 101.4 |
| food | 8.2% | 6.3% | 99.5 |

*All scores are statistically significant at $p < 0.001$ after a Benjamini Hochberg correction.

### 3.5.1 GPT-OSS-120B full text against title/abstract descriptive analysis

In terms of the topics of articles most affected by whether the input is full text or title/abstract, GPT-OSS-120B systematically scores articles on deep learning and other machine learning (relatively) higher if fed with their titles and abstracts than fed with their full texts. In contrast, it systematically scores articles about food quality lower if fed with their titles and abstracts than fed with their full texts (Figure 10). Gemma 3 27B differs by systematically scoring articles on randomised trials (relatively) higher if fed with their titles and abstracts than fed with their full texts. Moreover, it systematically scores articles about soil lower if fed with their titles and abstracts than fed with their full texts (Figure 11).

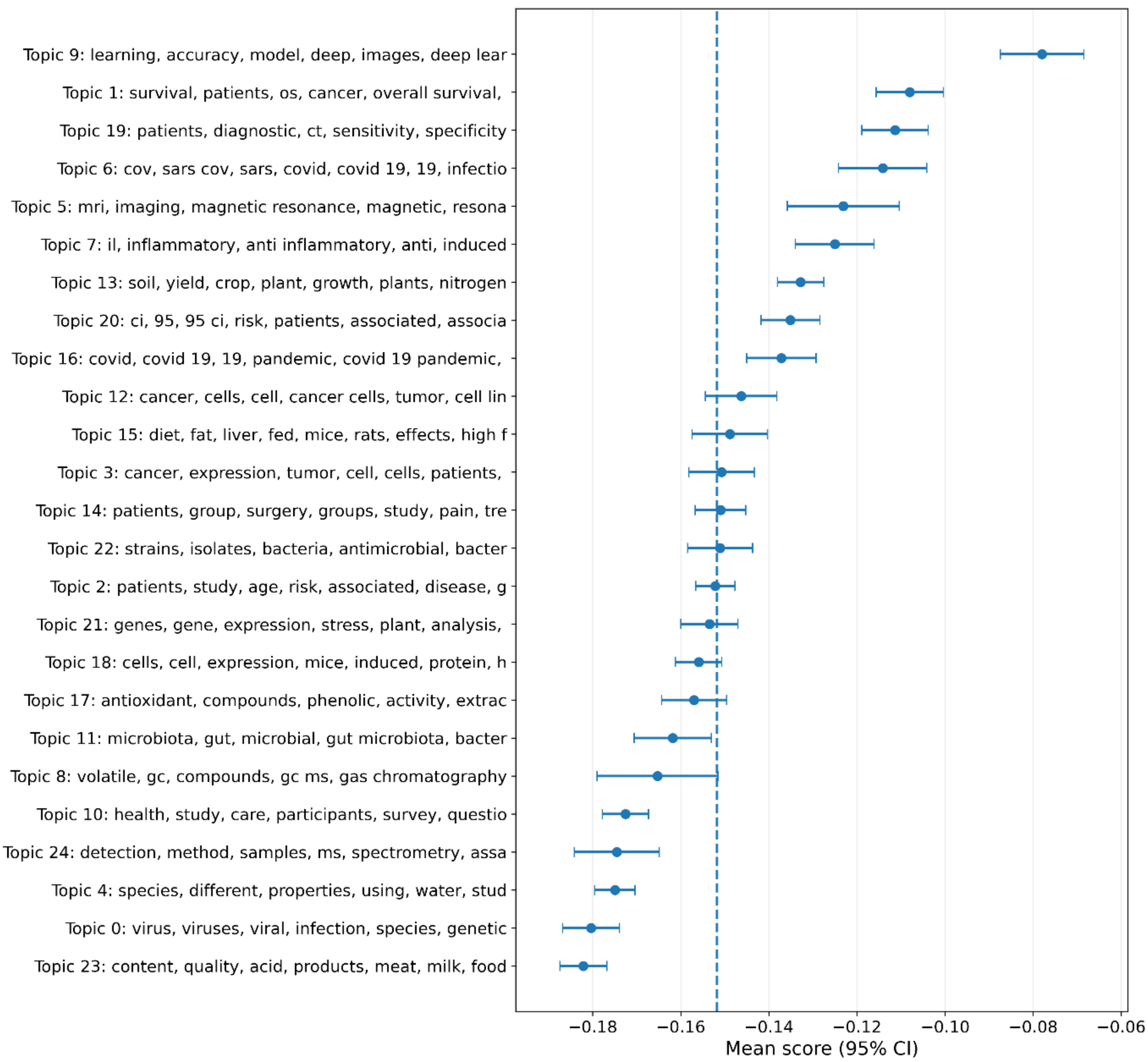

Figure 10. 25 article title/abstract clusters for all 15 life science journals combined and their mean GPT-OSS-120B title/abstract scores minus GPT-OSS-120B full-text scores. Similarity was judged by TF-IDF based on 1-3 word phrases. The dashed line is the overall mean difference.

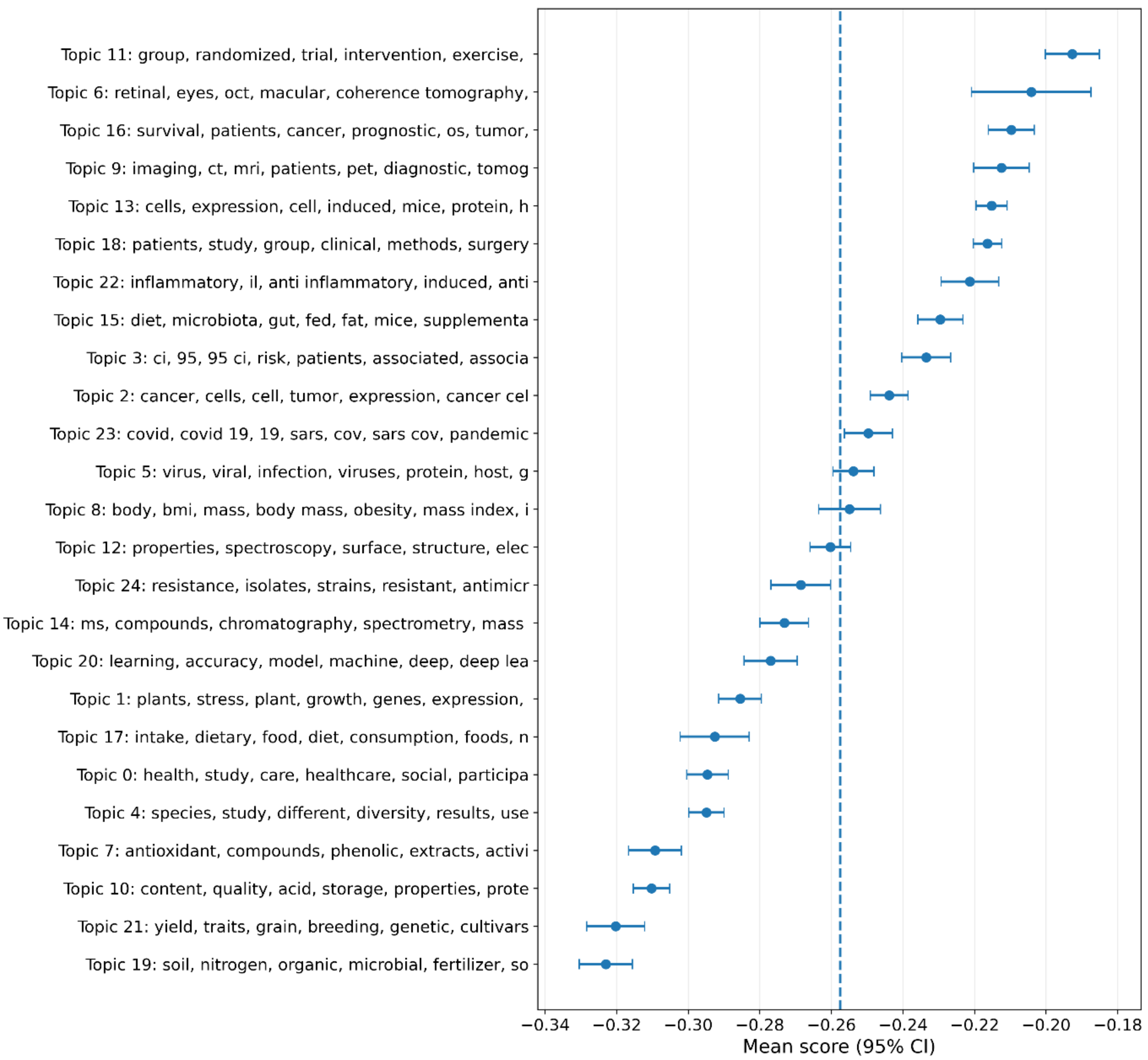

Figure 11. 25 article title/abstract clusters for all 15 life science journals combined and their mean Gemma 3 27B title/abstract scores minus Gemma 3 27B full-text scores. Similarity was judged by TF-IDF based on 1-3 word phrases. The dashed line is the overall mean difference.

# 4 Discussion

## *4.1 Limitations*

This study is primarily limited by the choice of journal articles and year range analysed since other publishers' journals may show different patterns. It is also restricted to the two LLMs investigated. Since these two LLMs had different patterns, other LLMs may also have different biases or none. In particular, larger LLMs seem to differ less between each other than do smaller LLMs (Kim et al., 2025), so the evidence of bias found here may be less evident or non-existent in large LLMs. The results also depend on the prompts used and the definition of research quality employed by the UK REF. Although its reliance on rigour, originality and significance are standard if not universal (Langfeldt et al., 2020), other quality criteria are also used internationally.

The interpretation is also limited by a lack of information about the quality of the articles analysed, such as from expert review scores. Thus, in the absence of this gold standard, the extent of bias in the LLMs is unclear and in the pairwise comparisons above (e.g., title/abstract vs. full text; Gemma 3 27B vs. GPT-OSS-120B) it is not clear whether one or both approaches are biased, even though at least one must be biased.

The statistical tests all violate the independence assumption to some extent due to authors potentially writing multiple related articles, or teams/departments publishing sets of related or similar articles. From a longer term perspective, the independence assumption is also violated by temporary research trends attracting high or low scores.

## 4.2 RQ1: Do any types of health and life science journal articles receive higher scores from LLMs within or across journals?

The word frequency chi-squared test results give strong statistical evidence that some types of health and life science journal articles receive higher scores from both GPT-OSS-120B and Gemma 3 27B within and across journals, giving a positive answer to RQ1. This does not prove LLM bias, however, since experts can give different average scores to journal articles in different fields, as occurred in REF2021 (Figure 3.2.2 of: Thelwall et al., 2022). For example, clinical research with patients, surveys, and education-related studies all seem to attract lower expert scores (Thelwall et al., 2023; Thelwall et al., 2025). Moreover, researchers may consider some topics to be inherently more important than others within their field (e.g., Hoekstra et al., 2023), and in some contexts, some topics can tend to get higher reviewer scores than others (Xu et al., 2026). Thus, the individual score disparities found in the current article do not necessarily indicate LLM bias since they may reflect topic-based expert score disparities.

A limitation of the cross-journal analysis is that individual journals in the set may be more selective than others, with their topic differences producing a second order effect. For example, some of the patterns above have an alternative explanation: if the journal *Viruses* is more selective, then this would explain the higher cross-journal average scores for articles mentioning viruses. Similarly, if *Healthcare* and *Nutrients* were less selective then this would explain the lower average scores for research with questionnaires (these two journals publish the most survey research in the set). It does not explain the within-journal differences, however.

## 4.3 RQ2: Do the types of health and life science journal articles that receive higher scores from LLMs differ between LLMs?

The word frequency chi-squared test results give strong statistical evidence for a positive answer to RQ2: there are statistically significant term differences between GPT-OSS-120B and Gemma 3 27B that probably reflect significant topic differences. For example, GPT-OSS-120B tends to give relatively higher scores to virus research and Gemma 3 27B tends to give relatively higher scores to machine learning research. Again, in the absence of a human expert gold standard to compare the LLMs against, either or both sets of LLM scores could be due to AI bias. As mentioned in the Introduction, it is known that scoring biases can vary between LLMs for other tasks (Bavaresco et al., 2025) so this difference is not surprising.

The reason for the relatively different average scores from GPT-OSS-120B and Gemma 3 27B for specific topics (e.g., viruses, machine learning) is unclear. OpenAI (2025a, 2025b) claims that the GPT-OSS series is trained on predominantly English text with a focus on Science, Technology, Engineering and Mathematics whereas Google (2025) only states that Gemma 3's training corpus is multilingual as well as multimodal. Thus, their training data may be substantially different, and this can lead to different scores such as through familiarity bias if one topic is covered in one corpus more than another (Stureborg et al., 2024). Another possibility is that the human feedback training stages differed to the extent that one LLM pays more attention to some aspects of the score than another (e.g., Gemma 3 considers rigour to be more important), biasing the results in favour of topics with articles that are strong in that dimension.

## *4.4 RQ3: Do the types of health and life science journal articles that receive higher scores from LLMs differ based on whether the LLM is fed with the full text or the title and abstract?*

The word frequency chi-squared test results give strong statistical evidence for a positive answer to RQ3. Most studies have used titles/abstracts as inputs for published journal article scoring because the results are not better if full text is entered, but it is now clear that this choice either adds a bias to the results, removes the bias from the results or (intuitively more likely) changes the nature of the results bias.

The reason for this disparity is again not obvious. The scores for full-text articles are generally higher, suggesting that the LLMs successfully mine the additional text for signals of rigour and originality and perhaps also significance. In this context, an obvious reason for disparities is that more complex research may not be able to adequately summarise its contributions within an abstract, so it benefits more from full-text processing. This is not a convincing explanation because other research has not shown increased LLM score accuracy from full texts, and the topics that benefit least from full text (e.g., machine learning) are complex. The opposite may therefore be the case: LLMs cannot "understand" (i.e., model well enough to respond appropriately) complex methods, and this causes them to guess that they are good based on extensive methods details irrespective of their actual quality.

## *4.5 Implications of between-topic LLM score disparities for practical research evaluations*

Although it was previously known that ChatGPT-4o mini, ChatGPT-4o and ChatGPT-5o mini are moderately accurate at ranking journal articles for research quality in most or all fields (Thelwall, 2026a) and that a range of medium sized LLMs are moderately accurate at ranking journal articles for research quality in the life and health sciences (Thelwall & Mohammadi, 2026), the current study definitively shows that at least one medium sized LLM and probably both Gemma 3 27B and GPT-OSS-120B give biased results with both reasonable inputs and also shows that the other gives biased results with at least one of its inputs (because their results differ), so both models and both input types are biased in some or all contexts.

The issue of bias within LLM results is different from that of their accuracy because bias suggests systematic unfairness and giving results that may change the nature of science, depending on the purpose. For instance, if LLM scores are used to evaluate job candidates' publications then this might result in fewer survey researchers and more virus researchers being appointed in the health sciences. Similarly, if LLM scores are used to help evaluate grant applicants then it might result in more virus research and less survey research being funded. Survey research is important or it would not be accepted for publishing in journals so this bias could be a long term problem. There are probably human biases already in both use cases, such as prestige bias and funding panel members being biased for or against some paradigms or approaches, so the importance of LLM biases should not be overstated. Nevertheless, it seems reasonable to at least check or monitor for the effects of LLM biases in important cases where they are used.

Perhaps the most influential potential use case for LLMs is in national research evaluation systems: to fully or partly replace expert judgment in existing systems such as those in the UK, New Zealand, or Italy, to fully or partly replace citation-based indicators in countries that use them, such as Poland, Sweden, and Slovakia, or to instigate national evaluation systems in countries that do not have them already, such as Bangladesh, Japan, or Mauritius. These can be more influential because, as in the UK, universities may be continually planning for the evaluation and making managerial decisions partly with the goal of improving their scores (Neyland & Milyaeva, 2025), so there is scope for LLM biases to have a wider influence on decision making. If universities know or suspect that LLMs will be used within the evaluation process then it would naturally lead them to mimic this in their own preparation procedures, using the same LLM if known and possible. Again, the extent of the systemic changes that might induce this should not be overstated because there are many pressures on researchers (e.g., teaching, funding, impact generation, publishing and field-based personal goals) and all existing evaluation systems have biases. Nevertheless, national evaluation systems are important, and it seems reasonable to be aware of potential problems caused by LLM bias before they are used and decide whether they are likely to be substantial enough to outweigh the cost, speed, or other advantages of using them.

# 5 Conclusions

The results give strong evidence that LLMs can have biases in terms of topics and perhaps also linguistic styles that tend to score above or below average for life and health science journal articles. The within-journal evidence is the least likely to be due to confounding factors, especially under the assumption that articles in a journal tend to have similar quality. The results also show that the nature of the bias changes if the score is based on the full text of an article rather than its title and abstract, even though the accuracy of the two inputs has previously been shown to be similar (with title/abstract input tending to align more closely with human experts). It is not clear which is the least biased LLM or input, however, since there is no human expert gold standard and any topic score disparity may closely match expert results (e.g., if both favour research on viruses).

Despite the above caveat, the results give the first statistically significant evidence of systematic biases in LLM research quality scores. This is a warning to those planning to use them to replace or supplement expert judgment. Of course, individual humans can also have biases, so if an LLM is used it will never clearly replace a demonstrably “correct” human score with its biased attempt. Nevertheless, using the same LLM to make all the decisions rather than a range of experts with (presumably) different biases seems likely to result in stronger overall decision biases. In addition, awareness of LLM use in decision making can lead to authors discovering the biases and switching topic to avoid them (e.g., giving up on survey research after reading the current article). This seems likely to skew science (further?) away from its ideal goals. Even if this risk is mitigated by varying the LLMs, the commonalities between Gemma 3 27B and GPT-OSS-120B suggest that cross-LLM biases may have a similar effect. This hints that LLMs should only be used to support and not replace human expert judgment for scoring published research, except perhaps for minor tasks, and if they are used in any significant role, steps should be taken to identify and mitigate bias.